\documentclass[journal]{IEEEtran}

\usepackage{amsmath,amsfonts}
\usepackage{algorithmic}
\usepackage{algorithm}
\usepackage{array}
\usepackage[caption=false,font=normalsize,labelfont=sf,textfont=sf]{subfig}
\usepackage{textcomp}
\usepackage{stfloats}
\usepackage{url}
\usepackage{verbatim}
\usepackage{graphicx}
\usepackage{cite}
\usepackage{float}  
\usepackage{lineno}
\usepackage{placeins}
\usepackage{amsmath}
\usepackage{gensymb}
\usepackage{hyperref}

\begin{document}

\title{Dual-Axis Beam-Steering OPA with purely Passive Phase Shifters }

\author{IEEE Publication Technology,~\IEEEmembership{Staff,~IEEE,}
        % <-this % stops a space

\author{
    Venus Kakdarvishi, Bowen Yu, and Yasha Yi %
    \thanks{Venus Kakdarvishi and Bowen Yu are with the Intelligent Optoelectronics Chip Laboratory, University of Michigan, Dearborn, MI 48128, USA. (e-mail: venuskd@umich.edu; bowenyu@umich.edu).}%
    \thanks{Yasha Yi is with the Intelligent Optoelectronics Chip Laboratory, University of Michigan, Dearborn, MI 48128, USA, and also with the Energy Institute, University of Michigan, Ann Arbor, MI 48109, USA. Corresponding author: Yasha Yi (e-mail: yashayi@umich.edu).}%
   
}

\thanks{This paper was produced by the IEEE Publication Technology Group. They are in Piscataway, NJ.}% <-this % stops a space
\thanks{Manuscript received April 19, 2021; revised August 16, 2021.}}


%\IEEEpubid{0000--0000/00\$00.00~\copyright~2024 IEEE}

\maketitle

\begin{abstract}

In this work, we present a multi-layer optical phased array (OPA) designed for dual-axis beam steering on a silicon (Si) platform, utilizing only wavelength tuning. Our design eliminates the need for grating couplers, commonly required for dual-axis beam steering, thereby reducing energy losses due to substrate leakage. It also features the unique capability of achieving a positive or negative phase slope profile. This three-dimensional architecture enhances output efficiency by emitting light from the device edge, providing greater flexibility and improved performance in beam steering. This approach opens up new possibilities for on-chip photonic systems, enabling faster, more accurate, and broader beam steering range. 
\end{abstract}


\section{Introduction}
\IEEEPARstart{o}ptical Phased Arrays (OPAs) have transformed the control and steering of light beams across various advanced applications, including free-space optical communication, holographic displays, imaging, and light detection and ranging (LiDAR).

OPA technologies leverage diverse mechanisms to achieve precise beam steering, with notable implementations including Liquid Crystal OPAs \cite{1035273}, Microelectromechanical Systems (MEMS) OPAs \cite{10.1117/12.2044197,mahjourian2024multimodalobjectdetectionusing}, Silicon Photonic OPAs (commonly referred to as solid-state OPAs) \cite{Wu:20 ,photonics11030243}, Electro-Optic Polymer OPAs \cite{10491204}, Acousto-Optic OPAs \cite{Akemann:15}, and Plasmonic OPAs \cite{zeng_all-plasmonic_2017}.

To achieve beam steering in solid-state OPAs—referred to as OPAs for simplicity hereafter—phase shifters are essential, which can be either active or passive. Active phase shifters, which often utilize the electro-optic \cite{8853396} or thermo-optic \cite{Akemann:15} effect to modulate the light's phase, typically suffer from high power consumption. Furthermore, OPAs must be capable of high-speed beam steering. Architectures based on active phase shifters employ lookup tables to determine the appropriate signal for each beam angle. In such setups, continuous steering necessitates a stabilization delay for the phase shifters at each step of the sweep. This significantly prolongs the sweep time, as it depends on both the number of angular steps and the necessary relaxation time for each step \cite{Yaacobi:14}. Therefore, developing a passive phase shifter solution is crucial for enhancing both efficiency and performance.

Traditional single-layer OPA configurations with an $M \times N$ array require $M \times N$ active phase shifters for two-dimensional (2D) beam steering. By incorporating grating couplers, 2D steering can be achieved with only $M$ phase shifters combined with wavelength tuning \cite{10024703, van2011two}. Despite this advantage, these designs face limitations in efficiently steering light across multiple dimensions, leading to increased complexity and potential inefficiencies \cite{lei2022effective, 10.1063/1.5000741}. Typically, these systems achieve one-dimensional steering through phased array principles, while wavelength tuning is used to control steering in the orthogonal direction, adding complexity to the design.

Emitter part in OPAs can be either end-fire (device edge) or grating couplers. Single-layer OPA configurations with grating emitters typically suffer from substrate leakage \cite{van2011two}, resulting in energy loss due to downward coupling from the grating structure \cite{VanAcoleyen:09}. This inefficiency restricts the effective steering of light beams, limiting the array’s functionality. Additionally, diffraction \cite{VanAcoleyen:09, 6044690} complicates OPA design optimization, as achieving desired beam convergence requires precise control over the grating period to ensure constructive interference. To mitigate crosstalk within OPAs, designs often employ strategies to suppress inter-waveguide interference, though these methods can increase device complexity. For instance, using waveguides with varying widths \cite{phare2018silicon, wu2018optical} facilitates the realization of an end-fire array, but applying this to a waveguide grating coupler configuration presents challenges, particularly in achieving a 2D converged beam. Furthermore, when light is emitted from the top of a grating device, reflections at various interfaces, such as the air-device boundary, can reduce the emitter's efficiency. To mitigate this issue, anti-reflection coatings are commonly applied \cite{Kwong:14}, requiring an additional step in the fabrication process. Achieving a narrower Full Width at Half Maximum (FWHM) for a smaller beam width typically requires a longer grating structure, but this spreads power along its length, reducing edge power density and overall emitter intensity \cite{VanAcoleyen:09}.

On the other hand, single-layer end-fire OPAs represent another type of emitter, producing a stripe-like (fan) beam \cite{10.1063/1.5000741, bhandari2022dispersive}, which limits steering capability to a single dimension.

As previously noted, most studies have employed active phase shifters combined with grating couplers for 2D beam steering. However, the works presented in \cite{10024703, Dostart:20} introduced the use of delay lines alongside grating couplers to implement fully passive phase shifters. Despite this advancement, the approach still faces limitations due to the inherent drawbacks of grating couplers. To overcome these challenges, we propose a novel OPA design that eliminates the need for grating couplers by utilizing passive phase shifters based solely on delay lines. Our design incorporates delay lines between arrays within each layer, as well as between corresponding arrays across layers, enabling precise control over phase distribution.

In our previous work \cite{Wu:23,Wu:24}, we introduced a multi-layer OPA design primarily aimed at enhancing input coupling efficiency and achieving beam steering in only one direction. Here, we leverage an additional advantage of the multi-layer design by incorporating delay lines between corresponding waveguides across different layers, introducing a novel method for 2D beam steering. We employed a non-uniform arrangement of emitters within each layer, while a uniform pitch in the vertical direction is maintained through cladding between layers. This design facilitates efficient beam steering in both orthogonal directions without the need for additional components. This design not only improves directivity but also produces a point-like output characteristic, significantly enhancing the efficiency of coupling to other optical components. This approach enhances system performance and versatility, providing a compact and scalable solution for advanced photonics applications.

\section{Design}

Beam steering in OPAs typically involves modulating the phase of the light emitted by each element. The steering angle in phased arrays is controlled by the phase difference and spacing (pitch) between each two adjacent emitters. When using a passive phase shifter based on delay lines, the interference pattern, analogous to that observed in a double-slit experiment, where \( n_{\text{eff}}(\lambda) \Delta L \) induces a shift in the fringes:

\begin{equation}
\scalebox{0.90}{$
d \sin \psi = m \lambda - n_{\text{eff}}(\lambda) \Delta L
$}
\label{eq:2}
\end{equation}

In this design, steering is achieved by tuning the wavelength of light, where $d$ is the pitch, $\psi$ is the steering angle, $\lambda$ is the wavelength of light, $\Delta L$ is the delay line difference, and $n_{\text{eff}}$ is the effective refractive index.

This OPA is designed for 2D beam steering by utilizing two distinct sets of delay lengths: one set within each layer ($\Delta L_1$), which introduces a phase difference ($\Delta \phi_1$) among elements in the same layer, and another set between corresponding waveguides across different layers ($\Delta L_2$), which introduces a phase shift ($\Delta \phi_2$) between layers. These two set delay lengths enable dual-axis control using only wavelength tuning. Specifically, $\Delta L_1$ facilitates steering in one direction, while $\Delta L_2$ enables steering in the orthogonal direction.

The differential phase shift $\Delta \phi$ caused by the delay lines is expressed as: $
\Delta \phi = \frac{2\pi \Delta L \, n_{\text{eff}}(\lambda)}{\lambda}$. Beam steering for one spot width can be calculated by differentiating this phase with respect to the wavelength, resulting in the wavelength step required for the phase shift: $\Delta \lambda = \frac{\lambda^2}{\Delta L \, n_{\text{eff}}(\lambda)}$. Applying both $\Delta L_1$ and $\Delta L_2$ allows the system to transition from 1D to full 2D steering. The equations for beam steering in the $y$ and $z$ directions are given by: $\psi_{y/z}(\lambda) = \sin^{-1} \left( \frac{\lambda}{d} - \frac{n_{\text{eff}}(\lambda) \Delta L_{1/2}}{d} \right)$, where $\Delta L_1$ and $\Delta L_2$ correspond to the delay line lengths for the respective axes. This configuration enables beam steering in one direction while simultaneously and repeatedly sweeping across a defined range in the orthogonal direction. To attain faster steering over a specific range, longer delay lines are required, this not only enables rapid steering but also allows for repeated sweeping within the range. This approach creates a fully available 2D field of view (FOV). This phenomenon arises from the periodic nature of the phase difference introduced by the delay lines. As the delay length ($\Delta L$) increases, the corresponding phase shift grows linearly. When this phase shift reaches an integer multiple of $2\pi$ (modulo $2\pi$), it aligns with the same diffraction order. This periodicity, which occurs at intervals of $\frac{\lambda}{n_{\text{eff}}(\lambda)}$, indicates that increasing $\Delta L$ leads to the recurrence of a specific steering range.

Each layer of the OPA incorporates an equal number of waveguides with non-uniform spacing to mitigate Side-Lobe Levels (SLL). To further optimize the pitch size and achieve minimal SLL along the horizontal, Particle Swarm Optimization (PSO) was implemented in \texttt{MATLAB}. The optimization process involved 64 adjustable parameters representing pitch size from $x = 0$, corresponding to the 8 emitters in each layer. The PSO algorithm, employing a swarm size of 200, converged after 436 iterations. To enhance solution refinement in the final stages, the \texttt{fmincon} function was integrated as a hybrid method alongside PSO.

Another advantage of the multi-layer design with a phase difference (\( \Delta \phi \)) between layers is its ability to enable positive or negative phase profile slopes by adjusting the phase gradient—specifically, the order of \( \Delta L \) between layers. This defines the beam steering direction, as constructive interference occurs along the phase gradient direction.

In this work, we use silicon (Si) as the waveguide material and silicon dioxide (\(\text{SiO}_2\)) as the cladding, taking advantage of their well-established benefits in photonic integrated circuits (PICs). Silicon’s high refractive index ($\sim 3.45$ at 1550 nm) ensures strong optical confinement, crucial for compact and efficient designs, while its transparency in the near-infrared region makes it ideal for optical communications. Moreover, Si’s compatibility with CMOS technology allows for cost-effective, scalable fabrication. Silicon dioxide, with a lower refractive index ($\sim1.44$ at 1550 nm), provides excellent cladding to keep light confined within the waveguide. The effective index (\(n_{\text{eff}}\)) of a single-mode silicon waveguide (500 nm width, 220 nm thickness) is approximately 2.5 at 1500 nm and 2.39 at 1600 nm. Across the wavelength range, \(n_{\text{eff}}\) for Si remains higher than that of \(\text{Si}_3\text{N}_4\), enabling nearly double the phase shift for the same delay length. As a result, the delay length can be halved with silicon, enhancing performance and compactness. Additionally, using a single-mode silicon (Si) waveguide allows us to achieve a larger envelope function. However, even with \(\text{Si}_3\text{N}_4\) emitters, silicon remains superior for phase shifting \cite{7831261, Wang:21, Pengfei:19}.

\section{Architecture and Simulation results} 
\begin{figure}[ht]
\centering
\begin{minipage}[t]{0.4\linewidth}
\centering
\subfloat[]{\includegraphics[width=\linewidth ,height=4cm, keepaspectratio]{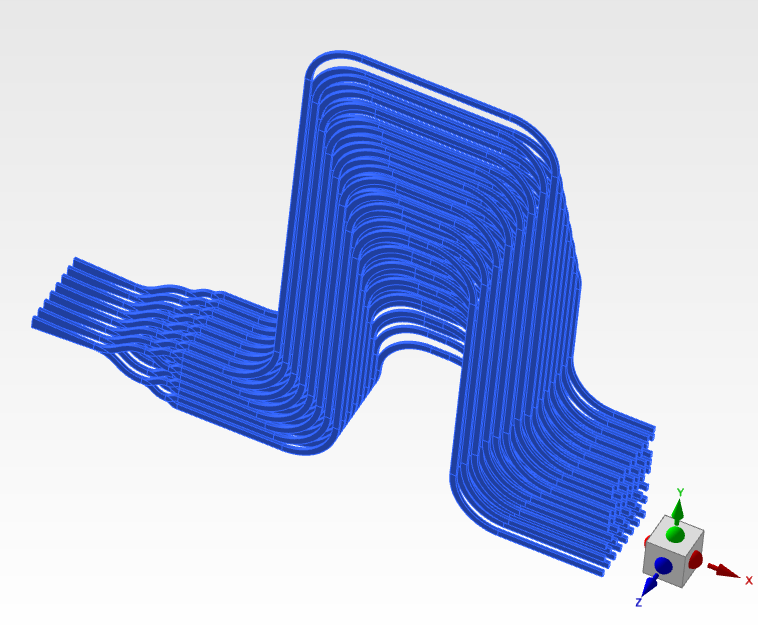}}
\end{minipage}%
\begin{minipage}[t]{0.6\linewidth}
\centering
\subfloat[]{\includegraphics[width=\linewidth ,height=4cm, keepaspectratio]{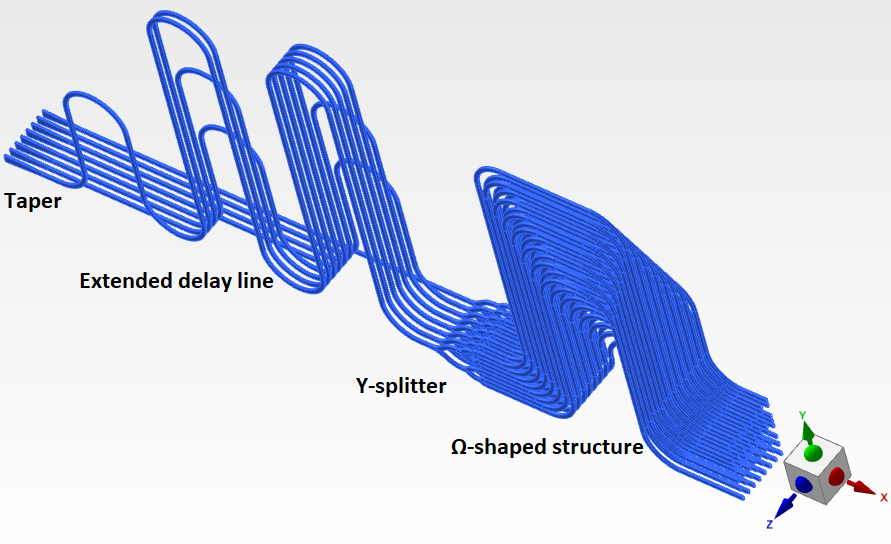}}
\end{minipage}
\begin{minipage}[t]{0.5\linewidth}
\centering
\subfloat[]{\includegraphics[width=\linewidth,height=4cm, keepaspectratio]{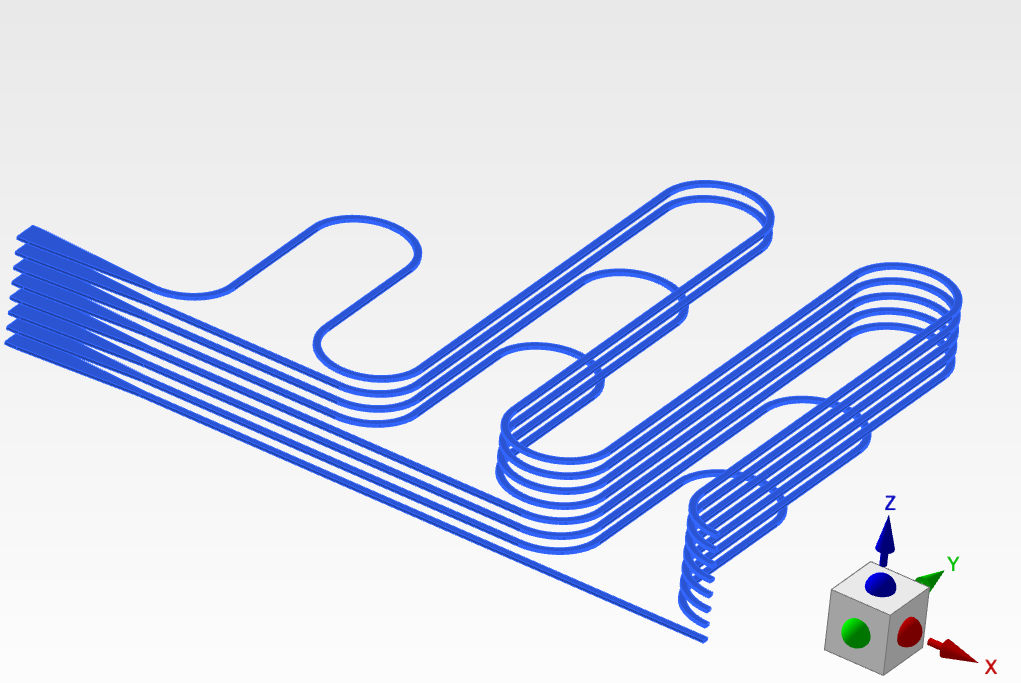}}
\end{minipage}%
\begin{minipage}[t]{0.5\linewidth}
\centering
\subfloat[]{\includegraphics[width=\linewidth,height=4cm, keepaspectratio]{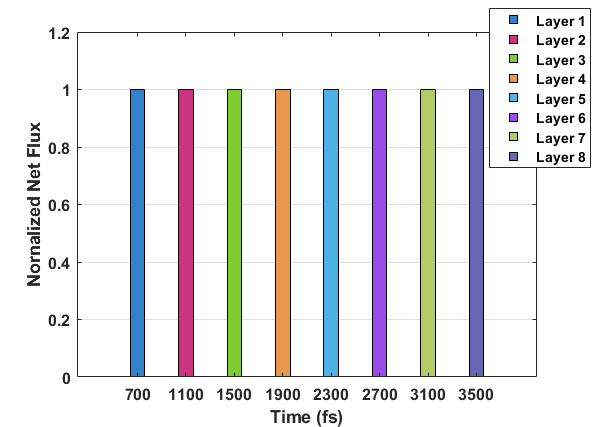}}
\end{minipage}
\captionsetup{font=small}
\caption{Schematic of 8-layer design, including: taper, Y-splitter, and $\Omega$-shaped structure with different values for $\Delta L_1$ and $\Delta L_2$. (b) Structure with an extended delay line (c) for $\Delta L_2$. (d) Normalized output flux versus time for each delay lines in different layers accroding to Fig. \ref{fig:1}(c).}
\label{fig:1}
\end{figure}

Our design schematic, as shown in Fig. \ref{fig:1}, features eight layers, each with eight waveguides.

In this design, the vertical angle is used for fast-axis sweeping, achieved by employing longer delay lines between layers, where we have a periodic pitch size. For the horizontal angle, which is used for slow-axis sweeping, smaller delay lines are implemented between waveguides within each layer, where aperiodic pitch sizes are applied between arrays. The longer delay line can either be introduced only in the \(\Omega\)-shaped part or in two stages: first, between the taper and the Y-splitter, and second, in an \(\Omega\)-shaped configuration. The second approach helps prevent the enlargement of the device, maintaining a compact footprint for larger delay lengths.

\FloatBarrier

\vspace{-5pt}

\FloatBarrier

Our design achieves a viewing angle of \(\sim78\) degrees in the vertical direction within a wavelength tuning range of \(\sim6\) nm, and a maximum viewing angle of 140 degrees in the horizontal direction with a 100 nm wavelength sweep, as shown in Fig.~\ref{fig:2}. The steering shows a linear relationship with wavelength, with slopes of $13^\circ$/nm, and $1.4^\circ$/nm for the vertical and horizontal directions, respectively (Figs.~\ref{fig:3}(a) and (b)). The beam profile remains consistent across wavelengths, with effective SLL suppression. The 3D far-field pattern (Fig.~\ref{fig:3}(d)) demonstrates the beam's focused and directive nature, ideal for 2D control. The tuning range from 1.5 to 1.6 µm provides a wide field of view (FoV), suitable for applications requiring broad directional scanning and fine angular resolution.

\begin{figure}[htbp]
\centering
\includegraphics[width=0.65\linewidth]{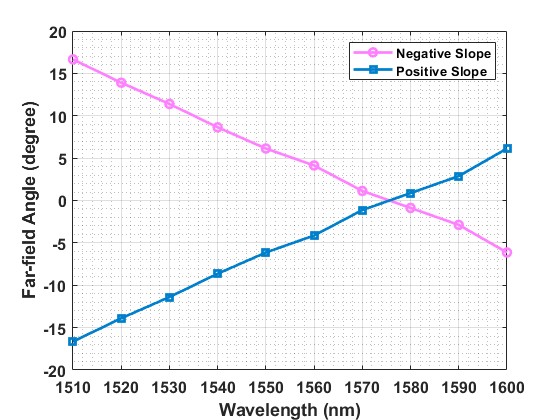}
\centering
\captionsetup{font=small}
\caption{Beam Steering in 8-Layer Designs: Reversing the phase slope mirrors steering, illustrating phase gradient impact on direction.}
\label{fig:4}
\end{figure}

In this design, a non-uniform pitch was employed for the waveguides in each layer to mitigate side lobe levels, with a PSO identifying the optimal configuration. These PSO-optimized values consistently achieved lower side-lobe levels compared to uniform configurations. The objective function used is the Side Lobe Suppression Ratio (SLSR) in decibels (dB) along the slow axis, with the main lobe centered at 0 degrees. The Maximum SLSR of -8.82 dB at 1550nm is observed . This non-uniform approach effectively reduced side lobes and minimized crosstalk while adhering to fabrication constraints. For inter-layer waveguides, a uniform pitch of 1.2~$\mu$m was employed, determined by the cladding thickness.
Another advantage of this design is the capability to achieve either negative or positive vertical phase profile slopes
, which is not possible with a grating-based emitter design. The order of \(\Delta L\) determines the phase gradient, which in turn sets the phase slope profile. This capability is illustrated schematically in Fig. \ref{fig:4}. 

\FloatBarrier
{
\setlength{\belowcaptionskip}{-40mm}
\begin{figure}[htbp]
\centering
\begin{minipage}[t]{1\linewidth}
\centering
\includegraphics[width=\linewidth]{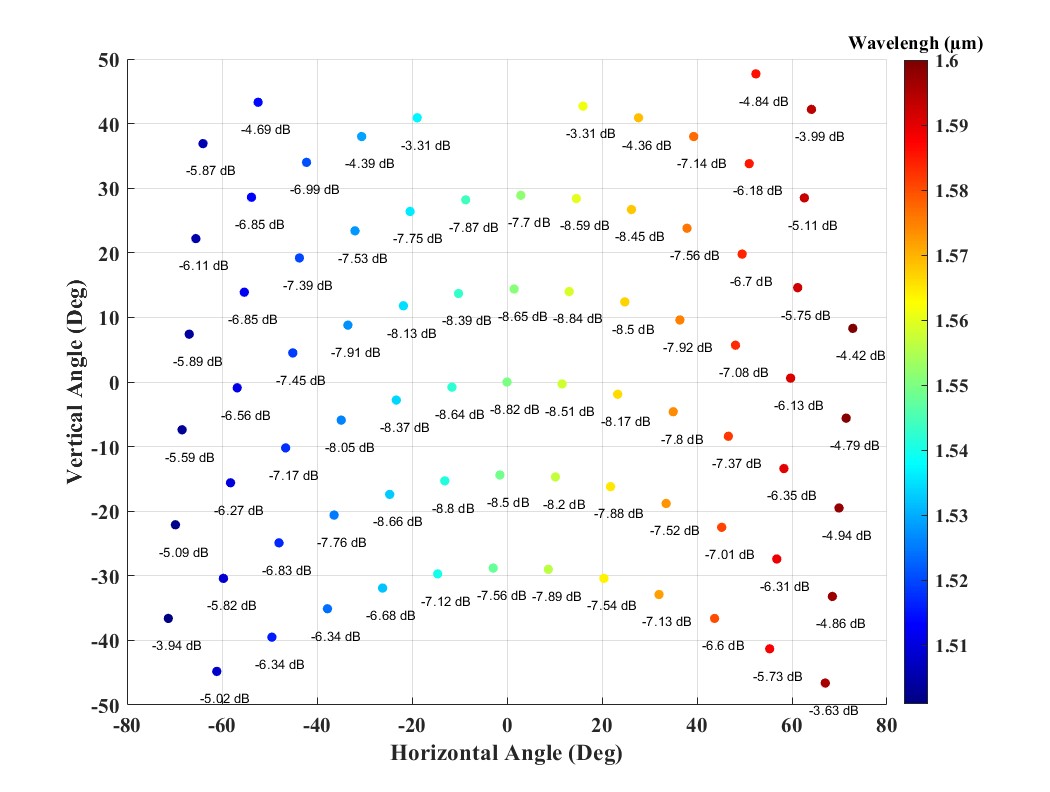}
\end{minipage}
\captionsetup{font=small}
\caption{Steering performance in two orthogonal directions. Each data point represents the main lobe for a specific wavelength, spanning from 1500 nm to 1600 nm. For each main lobe, the wavelength and corresponding side-lobe level are shown.}

\label{fig:2}
\end{figure}
}

\FloatBarrier

{
\setlength{\belowcaptionskip}{-30mm}

\begin{figure}[htbp]
\centering
\begin{minipage}[t]{0.5\linewidth}
\centering
\subfloat[]{\includegraphics[width=\linewidth]{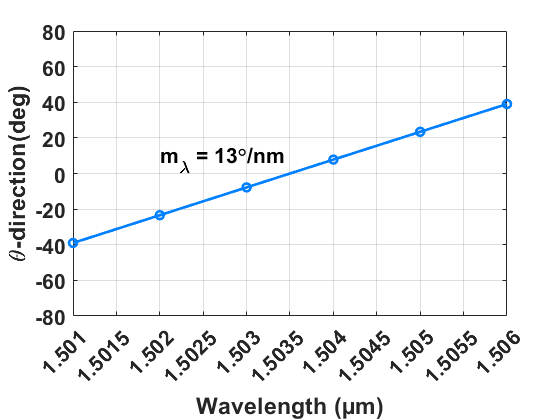}}
\end{minipage}%
\begin{minipage}[t]{0.5\linewidth}
\centering
\subfloat[]{\includegraphics[width=\linewidth]{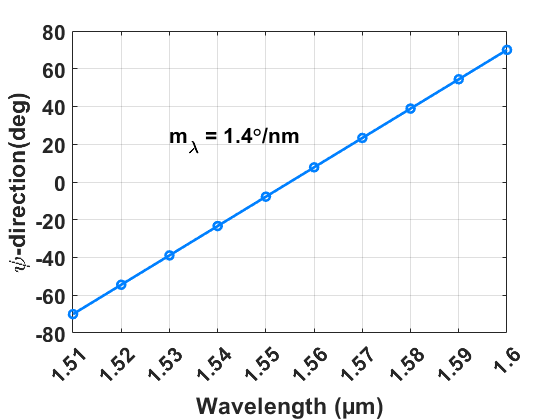}}
\end{minipage}
\begin{minipage}[t]{0.45\linewidth}
\centering
\subfloat[]{\includegraphics[width=\linewidth]{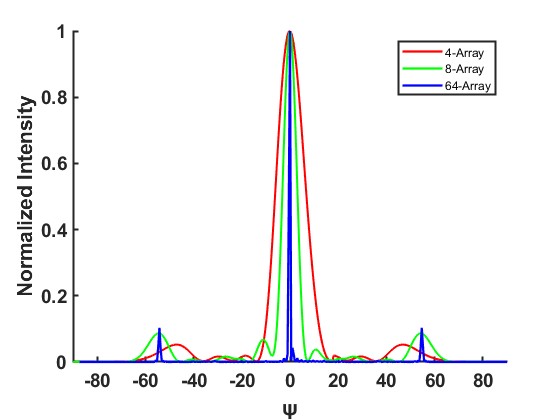}}
\end{minipage}%
\begin{minipage}[t]{0.55\linewidth}
\centering
\subfloat[]{\includegraphics[width=\linewidth]{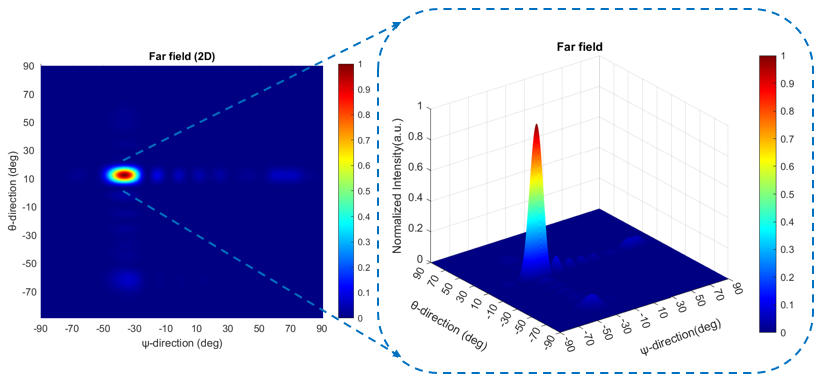}}
\end{minipage}
\label{fig:}
\captionsetup{font=small}
\caption{(a), (b) Linear fitting of the wavelength tuning efficiency (m$_{\lambda}$) in the $\theta$ (vertical) and $\psi$ (horizontal) directions, respectively. (c) FWHM for different numbers of waveguides. (d) Each point in Fig.\ref{fig:2} corresponds to a far-field point, as illustrated in this figure.}
\label{fig:3}
\end{figure}
}

\FloatBarrier

\section{Method} 
For all simulations, the refractive indices of Si and SiO\textsubscript{2} were chosen from the Lumerical material library (Ansys Inc.) A finite-difference eigenmode (FDE) solver was used to identify the effective refractive indices and mode field profiles for the theoretical calculations. All other simulations were conducted using a three-dimensional (3D) finite-difference time-domain (FDTD) method from Ansys Lumerical, as well as Omnisim software.

\FloatBarrier
\FloatBarrier

\FloatBarrier

\vspace{-6mm}

\section{Conclusion} 
We present a multi-layer optical phased array (OPA) with non-uniform pitch sizes between arrays in each layer and a uniform arrangement between layers, designed for dual-axis beam steering on a Si/SiO$_{2}$ platform. By integrating passive phase shifters based on delay lines across multiple layers, our OPA achieves wide-angle steering in two orthogonal directions, resulting in a large field of view (FOV). Numerical simulations validate the effectiveness of this approach. By eliminating the need for grating couplers, our design simplifies the system architecture and beam steering control. Furthermore, our innovative method enables the manipulation of phase profile slope, allowing for either positive or negative phase slopes through delay line gradients between layers.

\bibliographystyle{IEEEtran}
\bibliography{ref}

\end{document}